\documentclass[preprint2]{aastex}

\usepackage{ulem}

\def\Ms{$M_{\odot}$}
\def\psr{J1717$+$4308A}

\shorttitle{The FAST Discovery of a New Pulsar in the M92}
\shortauthors{Pan et al.}

\begin{document}

\title{The FAST discovery of an Eclipsing Binary Millisecond Pulsar in the Globular Cluster M92 (NGC 6341)}

\author{Zhichen Pan\altaffilmark{1,2}, Scott M. Ransom\altaffilmark{3}, Duncan R. Lorimer\altaffilmark{4,5}, William Fiore\altaffilmark{4,5}, Lei Qian\altaffilmark{1,2}, Lin Wang\altaffilmark{1,2,6,7}, Benjamin W. Stappers\altaffilmark{7}, George Hobbs\altaffilmark{8}, Weiwei Zhu\altaffilmark{1,2}, Youling Yue\altaffilmark{1,2}, Pei Wang\altaffilmark{1,2}, Jiguang Lu\altaffilmark{1,2}, Kuo Liu\altaffilmark{1,2,9}, Bo Peng\altaffilmark{1,2}, Lei Zhang\altaffilmark{1,6,8}, Di Li\altaffilmark{1,2,6,*}}

\affil{$^1$National Astronomical Observatories, Chinese Academy of Sciences, 20A Datun Road, Chaoyang District, Beijing, 100101, China}
\email{panzc@bao.ac.cn, dili@bao.ac.cn}
\affil{$^2$CAS Key Laboratory of FAST, National Astronomical Observatories, Chinese Academy of Sciences, Beijing 100101, China}
\affil{$^3$National Radio Astronomy Observatory, Charlottesville, VA 22903, USA}
\affil{$^4$Department of Physics and Astronomy, West Virginia University, Morgantown, WV 26506, USA}
\affil{$^5$Center for Gravitational Waves and Cosmology, West Virginia University, Chestnut Ridge Research Building, Morgantown, WV 26505}
\affil{$^6$College of Astronomy and Space Sciences, University of Chinese Academy of Sciences, Beijing 100049, China}
\affil{$^7$Jodrell Bank Centre for Astrophysics, School of Physics and Astronomy, The University of Manchester, Manchester, M13 9PL, UK}
\affil{$^8$CSIRO Astronomy and Space Science, PO Box 76, Epping, NSW 1710, Australia}
\affil{$^9$Max-Plank-Institut f{\"u}r Radioastronomie, Auf dem H{\"u}gel 69, Bonn, D-53121 , Germany}

\begin{abstract}
  We report the discovery of an eclipsing binary millisecond pulsar in the globular cluster M92 (NGC6341) with the Five-hundred-meter Aperture Spherical radio Telescope (FAST).
  PSR~\psr\, or M92A, has a pulse frequency of 316.5~Hz (3.16~ms) and a dispersion measure of 35.45~pc~cm$^{-3}$.
  The pulsar is a member of a binary system with an orbital period of 0.20~days around a low-mass companion which has a median mass of $\sim$0.18~\Ms.
  From observations so far, at least two eclipsing events have been observed in each orbit.
  The longer one lasted for $\rm \sim$5000~s in the orbital phase range 0.1--0.5.
  The other lasted for $\rm \sim$500~s and occurred between 1000--2000~s before or after the longer eclipsing event.
  The lengths of these two eclipsing events also change.
  These properties suggest that \psr\ is a ``red-back'' system with a low-mass main sequence or sub-giant companion.
  Timing observations of the pulsar and further searches of the data for additional pulsars are ongoing.
\end{abstract}


\section{Introduction}

There are currently 156 pulsars known in 29 globular clusters\footnote{http://www.naic.edu/$\sim$pfreire/GCpsr.html} (GCs).
One of the highlights from previous discoveries in globular clusters is the large population of eclipsing binary pulsars, 
which includes the fastest spinning millisecond pulsar, Terzan 5ad (Hessels et al. 2006). 
This object belongs to a subset of eclipsing systems known as "red-backs" (see, e.g., Crawford et al. 2013; Roberts 2013 and references therein)
Defining characteristics of red-backs include:
(1) the orbital eccentricity being close to zero (noting that the largest eccentricity measured to date if $<10^{-4}$ for PSR~J1740$-$5340 in NGC6397; D'Amico et al.~2001);
(2) a low-mass (typically 0.2 to 0.7\,M$_\odot$) main sequence companion star;
(3) short orbital periods (among the 12 known GC red-back systems only three, Terzan 5P, Terzan 5ad and NGC~6397A, have orbital periods longer than one day);
(4) ionized material from the companion star surrounding the pulsar (Bellm et al. 2016 and references therein);
(5) X-ray counterpart identifications x-ray (e.g., PSR~J1824$-$2452I or M28I, Pallanca et al. 2013; Papitto et al. 2013) or optical (e.g., NGC6397A, Ferraro et al. 2001).
All the GC red-backs are eclipsing binaries which even have large eclisping durations than other eclipsing pulsars.  
For example, 47\,Tuc V and W, typically have 50\% and 30\% durations, respectively (Ridolfi et al. 2016), and Terzan 5A is at $\sim$30\% typically and sometime eclipsing for the whole orbit (Bilous et al. 2019).
Sometime, at both eclipse ingress and egress, the pulses usually exhibit excess propagation delays as the pulses pass through the atmosphere of the companion.
These delays can be used to estimate the electron column density through the orbit (e.g., as seen in PSR~J1701$-$3006B or M62B; Possenti et al.~2003).
As with other binary pulsars, the orbital inclination can be calculated assuming that the companion is a low-mass main sequence star or a white dwarf (e.g., NGC6397A, D'Amico et al.~2001).

Pulsars in GCs are prime targets for deep radio searches with large single dishes.
Located outside of Arecibo sky and with a reasonable distance, the GC M92 (NGC6341) was selected as one of the highest priority targets for an early science survey with the
Five-hundred-meter Aperture Spherical radio Telescope (FAST; Nan et al.~2011 and Jiang et al.~2019).
M92 is a metal-poor (e.g., Roederer \& Sneden 2011) GC with a distance of $8.3 \pm 1.6$~kpc (Rees 1992).
It has been studied at optical wavelengths (see, e.g., the main-sequence color-magnitude diagram obtained by Heasley \& Christian 1986).
M92 has also been examined in the comparison to other clusters (e.g., with M13 and 47~Tuc, Cathey 1974, with M15 and NGC5466, Buonanno, Corsi \& Fusi Pecci, 1985),
and in X-ray (e.g., Lu et al. 2011; Fox et al. 1996).
According to an empirical relation (Hui, Cheng \& Taam 2010; Tuck \& Lorimer 2013) contingent upon the stellar interaction rate,
Zhang et al.~(2016) estimated the population size of M92 pulsars to be about 13, ranking the 5$^{\rm th}$ among all clusters visible to FAST.

In this paper, we present the FAST discovery and initial timing observations of an eclipsing binary millisecond pulsar in M92, PSR~\psr\ (hereafter M92A).
In Section 2, we describe the discovery observation, along with follow-up observations with FAST and the Green Bank Telescope (GBT).
We discuss the implications of this discovery in Section 3, and present our conclusions in Section 4.

\section{Observation, Data Reduction, and Timing Analysis}

The initial search observation of M92 was carried out on 2017 October 9$^{th}$ as part of a pilot survey for pulsars in GCs with FAST.
An ultra-wide-band receiver system was used that covered 270 to 1620\,MHz with a $\sim$60\,K system temperature.
The two polarization channels were sampled and digitized using 8-bit precision every 200\,$\mu$s.
The total integration time was 30\,min.
The signal processor directly sampled the data in the 0 to 2048\,MHz band, corresponding to 8192 channels in total and a channel bandwidth of 0.25\,MHz.
To avoid large time smearing at low frequency and smaller field-of-view at high frequencies,
we found an optimal band between 500 and 700\,MHz for subsequent processing.
In this band, the beam size (the full width half maximum) is between 4.9 and 6.9 arcminutes.
The diameters of GCs, including M92, usually are larger than 10 arcminutes,
while most of them have a much smaller core radius and half-light radius.
For M92, its core radius and half-light radius are 0.26 and 1.02 arcminutes, respectively (Harris 1996).
Our beam therefore covers the region of M92 in which we would expect to detect pulsars.

The searching was done with PRESTO\footnote{\rm https://www.cv.nrao.edu/~sransom/presto}
(PulsaR Exploration and Search Toolkit, Ransom 2001; Ransom, Eikenberry \& Middleditch 2002; Ransom, Cordes \& Eikenberry 2003).
Assuming a distance to M92 of 8.3~kpc (Rees 1992),
the dispersion measure (DM) predicted by the NE2001 model is 44 pc~cm$^{-3}$ (Cordes \& Lazio 2002; Cordes \& Lazio 2003)
and for the YMW16 model it is 35 pc cm$^{-3}$ (Yao, Manchester \& Wang 2017).
Taking the average value between these two models, the time smearing in one channel for a signal with a DM of 39.5~pc~cm$^{-3}$ is 0.2--0.7~ms across the whole band.
Given these predictions, we conservatively adopted a DM range of 0--100~pc~cm$^{-3}$ in the search.
To retain sensitivity to binary pulsars, we performed 1-D acceleration searches with a range of
$\pm 231$~m~s$^{-2}$ (assuming a 200~Hz signal) or $\pm 93$~m~s$^{-2}$ for a 500 Hz signal using the PRESTO routine {\tt accelsearch}.
Details of these calculations can be found in Ransom et al.~(2001).
We found a signal with a period $P=3.16$~ms with an acceleration of $-$15.5~m~s$^{-2}$,
and DM of $35.45$~pc~cm$^{-3}$ was found, very close to the value predicted from the YMW16 electron density model.
The measured signal-to-noise ratio was 51.
We note that without the acceleration search, the pulsar would have been undetectable.

In mid-2018, we started monitoring observations at FAST using the 19-beam L-band receiver.
This system covers a band of 1.05---1.45 GHz with a beam size of about 3 arcminutes and a system temperature on cold sky of about 24~K (Jiang et al. 2020).
For these observations, we sampled the data from two polarizations with 8-bit precision from the central beam every 49.152~$\mu$s with 4096 channels each of width 0.122~MHz.
So far, a total of 13 observations have been carried out in the 421-day interval between August 2018 (MJD 58351) and October 2019 (MJD 58711).
Depending on the FAST commissioning schedule, observations lasted between 0.5 to 5 hours.
For each observation, we dedispersed the data to a DM value of 35.45~pc~cm$^{-3}$,
searched for the signal, and obtained Times of Arrival (TOAs) for timing.
The first template used for TOA generation was made from data taken on 2018 September 18 using the accelerated spin parameters determined from a search of that day's data.
After determining an initial orbital solution using all the TOAs from 2018,
we folded the longest observation with that ephemeris and obtained an improved pulse profile.
A gaussian-based template, fitted to that pulse profile, was used to determine all of the TOAs for all the other observations with FAST 19-beam L-band receiver.
With the exception of two observations where the pulsar was undetected due to eclipses,
we obtained 1166 TOAs from eleven observations.
We normally obtained 128 TOAs from every observation, that is to say, every TOA was obtained from $\rm \sim$30--60~s observation.

\begin{figure*}
  \centering
    \includegraphics[width=160mm]{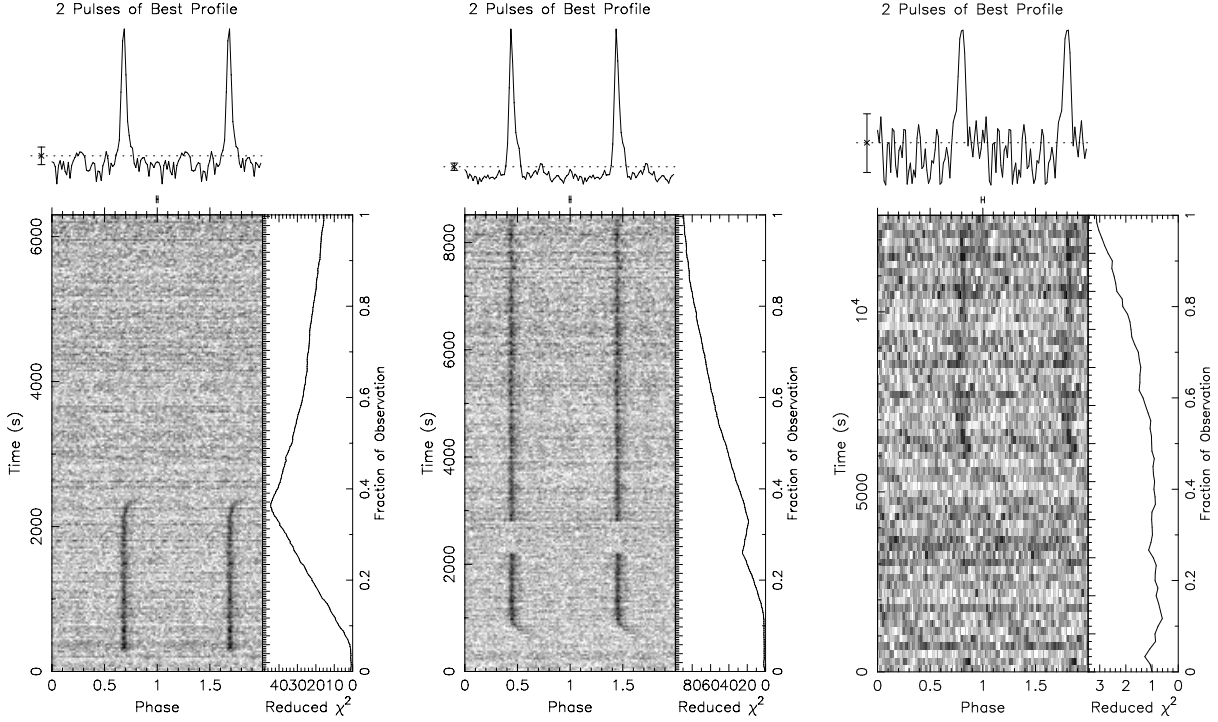}
      \caption{M92A eclipses.
      Left: folding result of FAST data taken on 2018 September 13$^{\rm th}$.
      The short eclipse event happened two thousand seconds before the longest eclipse (at the beginning of the observation).
      Middle: folding result of FAST data taken on 2018 October 2$^{\rm nd}$.
      The short eclipse event happened approximately one thousand seconds after the longest eclipse.
      Right: folding of GBT data taken on 2018 June 29$^{\rm th}$ to 30$^{\rm th}$,
      showing the huge sensitivity difference between GBT and FAST. }
  \label{eclipsings}
\end{figure*}

We also observed the pulsar with the GBT on 2018 June 29 and 30 for 3.5 hours (across two days).
These observations used the 820~MHz receiver and acquired data with the Green Bank Ultimate Pulsar Processing Instrument (GUPPI; DuPlain et al.~2008).
The GUPPI data were sampled with 8-bit precision every 81.92~$\mu$s over a bandwidth of 200~MHz split into 4096 contiguous channels.
We searched the data using PRESTO and the ``jerk-search'' functionality of \texttt{accelsearch} (Andersen \& Ransom 2018),
we obtained a candidate of a similar period to that seen in the FAST data.
While we were able to detect the pulsar,
due to the low signal-to-noise ratios of the detections and lack of detections during times where the pulsar was eclipsed,
we were unable to use them to obtain a solution for the binary orbit.
Later on, after we had successfully obtained a phase-connected timing solution from FAST data alone,
we folded the GBT data successfully.
The folded GBT observation is shown in the right panels of Fig.~\ref{eclipsings} alongside a FAST observation and highlights the stark contrast in sensitivity of the two telescopes.

\begin{figure*}
    \includegraphics[width=160mm]{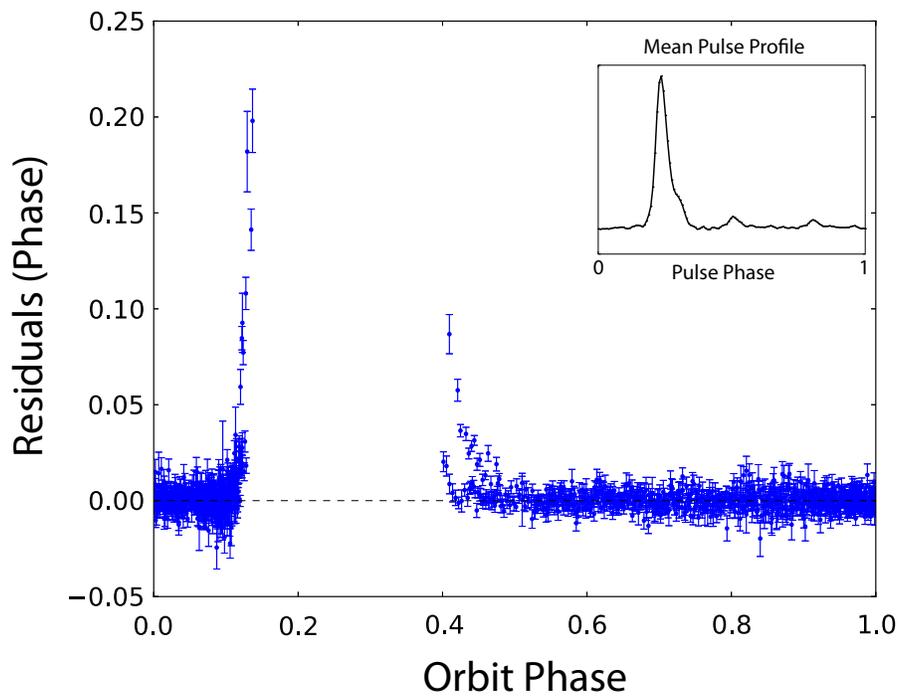}
      \caption{Timing residuals as a function of orbital phase and the mean pulse profile (upper right). }
  \label{timing_residuals}
\end{figure*}

Obtaining timing solutions for red-back systems is notoriously difficult.
We used the TEMPO\footnote{http://tempo.sourceforge.net} and TEMPO2 (see, e.g., Hobbs, Edwards \& Manchester 2006) software packages to carry out the timing analysis of the FAST TOAs.
With only 13 observations over a one$-$year period, including a gap of 9 months,
the timing solution is likely not completely phase connected over the full year.
We do not have a sufficient number of observing epochs to break degeneracies between the positional and spin-down parameters.
However, as the orbital period is short, we are confident in the orbital parameters and the timing residuals as a function of orbital phase.
These were obtained by fitting from phase offsets between each of the observations and assuming that the pulsar is at the center of the cluster (Harris 1996; Harris 2010).
In TEMPO, we put jumps between every observation, set the pulse frequency derivative and orbital eccentricity to be zero,
and used the cluster center as the position for the pulsar for obtaining timing solution.
In TEMPO2, we were able to obtain ''a tentative'' coherent timing solution across the entire data span
(without fitting for offsets between each observation) with a pulse frequency of 316.4836857767(2) Hz and no spin-down.
The results from TEMPO or TEMPO2 generally agree.
In Table \ref{timing_parameters} and Figure \ref{timing_residuals} we show the timing parameters and the timing residuals from obtained using TEMPO.
We are continuing to observe the pulsar and a ''final'' coherent timing solution will be published elsewhere,
along with possible multiwave-length counterparts and better constraints on the eclipsing medium from the companion star.

\begin{table*}
\centering
\caption{Parameters for PSR~J1717+4308A, M92A.}\label{timing_parameters}
\begin{footnotesize}\begin{tabular}{ll}\hline\hline
\multicolumn{2}{c}{Timing observation summary} \\
\hline
MJD range\dotfill & 58351---58711 \\
Data span (days)\dotfill & 421 \\
Number of TOAs\dotfill & 1166 \\
RMS timing residual ($\mu$s)\dotfill & 13.4 \\
\hline
\multicolumn{2}{c}{Measured quantities} \\
\hline
Pulse frequency (Hz)\dotfill & 316.4836857(3) \\
Orbital Period, P$_b$\dotfill & 0.2008678775(8) \\
Epoch of passage at Periastron, T$_0$ (MJD)\dotfill & 58353.5490817(3) \\
Projected Semi-major Axis, $\chi_p$ (lt-s)\dotfill & 0.398703(2) \\
\hline
\multicolumn{2}{c}{Set quantities} \\
\hline
Right ascension, RA (hh:mm:ss)\dotfill &  17:17:07.39 (J2000) \\
Declination, DEC (dd:mm:ss)\dotfill & +43:08:09.4 (J2000) \\
Reference epoch (MJD)\dotfill & 53600 \\
Dispersion measure, DM (cm$^{-3}$~pc)\dotfill & 35.45 \\
Pulse frequency derivative (s$^{-2}$)\dotfill & 0 \\
Orbital eccentricity, $e$\dotfill & 0 \\
Longitude of periastron, $\omega$ (deg)\dotfill & 0 \\
\hline
\multicolumn{2}{c}{Timing model assumptions} \\
\hline
Clock correction procedure\dotfill & TT(BIPM) \\
Solar System ephemeris model\dotfill & DE436 \\
Binary model\dotfill & BT \\
\hline
\end{tabular}
\end{footnotesize}
\end{table*}

\section{Discussion}

We believe that the pulsar is associated with M92 as:
(1) the measured DM value is close the model predictions;
(2) the properties of the pulsar (such as it being a millisecond pulsar, in a compact orbit, with a low mass companion),
are similar to many other GC pulsars Of course, a definitive association would require to discovery of another pulsar in the cluster with similar DM.
Searches are ongoing, but no second pulsar has yet been discovered.


M92A seems to be a new GC red-back pulsar with highly variable eclipses and occasional ``mini'' eclipses at non-standard orbital phases in each orbit.
The longest eclipsing event lasts for 5000\,s, which corresponds to $\sim$40\% of the orbit.
We note that the eclipsing event in M92A has a longer duration than an event purely caused by the physical extent of the companion star and is therefore caused by an ionized ``wind''.
A small eclipsing event has been seen in four time of our eleven observations.
The event lasts for 1000--2000~s (6--12\% of the orbital period) before or after this ``normal'' eclipsing event.
These minor eclipsing events are likely caused by anisotropic and clumpy ionized gas in the system.
The length of these two eclipsing events change from orbit to orbit.
Fig.~\ref{eclipsings} shows two observations as examples.

Bilous et al.~(2019) also reported that the intensity of the individual pulses from Terzan 5A at both eclipse ingress and egress
sometimes became up to 40 times higher than average likely due to plasma lensing.
Such a phenomena also happens in other eclipsing systems such as black widows.
Main et al.~(2018) reported plasma lensing in the original black widow, B1957+20, where the pulsar signal can sometimes be enhanced by factors of up to 70--80 at $\rm \sim$300~MHz.
Assuming that FAST has a system temperature of 20~K and gain of 16~K~Jy$^{-1}$,
then a single pulse from a pulsar with a width 1~ms and a peak flux of 18~mJy will be detected with a signal-to-noise ratio of 6 by FAST.
If plasma lensing can increase the intensity by a factor of 50, with FAST we are able to detect single pulses from millisecond pulsars down to a peak flux density of $\sim$0.36~mJy.
These results suggest that single pulse searches should also be used for GC pulsar observations to potentially aid in finding new red-backs and black widows.
We have applied a single pulse search algorithm to our M92 data, but no single pulses with signal-to-noise ratio larger than 6 have yet been detected.

The timing residuals near eclipse ingress and egress also show significant dispersive delays (see Fig.~2).
The maximum delay estimated from the data is approximately 0.6\,ms, corresponding to a DM excess of 0.3\,pc\,cm$^{-3}$.
This is similar to some other GC red-backs (e.g., less than 1 pc cm$^{-3}$ for NGC6397A, D'Amico et al.~2001) and significantly smaller than Terzan 5A
(see the millisecond-duration delays in Bilous et al.~2019).

\begin{table*}[htpb]
\hspace{-2em}
\centering
\caption{Timing observations of M92A and eclipsing events. Only eclipses in which the entire eclipse duration has been observed were measured.}
\label{eclipsing_all}
\begin{tabular}{cccc}
\hline
Date              & Observation Length  &  Eclipsing Length  &  Comments  \\
(YYYYMMDD)        &   (hours)           &       (s)          &            \\
\hline
20180821          & 3                   &        387         & Short, before the longer one\\
                  &                     &                    & Orbit phase: 0.08-0.10 \\
                  &                     &                    & Observation ended with eclipsing \\
\hline
20180823          & 5                   &                    & Observation started and ended with eclipsing \\
\hline
20180913          & 2                   &                    & Observation started and ended with eclipsing \\
\hline
20180914         & 2                   &                    & Observation ended with eclipsing \\
\hline
20180918          & 1                   &                    & No eclipsing observed \\
\hline
20181002          & 5                   &   5255             & A full observation of the longer one\\
                  &                     &                    & Orbit phase: 0.14-0.45\\
                  &                     &  591               & Short after the longer one\\
                  &                     &                    & Orbit phase: 0.52-0.56\\
\hline
20181003          & 5                   &  4469              & A full observation of the longer one\\
                  &                     &                    & Orbit phase: 0.15-0.41\\
                  &                     & 627                & Short, after the longer one\\
                  &                     &                    & Orbit phase: 0.50-0.54\\
\hline
20181004          & 0.5                 &                    & Full eclipsing \\
\hline
20180621          & 0.5                 &                    & Full eclipsing \\
\hline
20180625          & 1                   &                    & No eclipsing observed \\
\hline
20180627          & 0.5                 &                    & Observation ended with eclipsing \\
\hline
\end{tabular}
\end{table*}

\section{Conclusion}

A red-back eclipsing binary millisecond pulsar, PSR J1717+4308A or M92A,
with a pulse frequency of 316.48~Hz (corresponding to a pulse period of 3.1597~ms) and orbital period of 0.2~days was discovered in GC M92 with FAST.
This brings the number of GCs with pulsars to 30 and the total number of GC pulsars published to 157.
The pulsar's DM of 35.45~cm$^{-3}$~pc is very close to that predicted by the YMW Galactic electron density model.
Based on its orbital characteristics, the orbiting companion is most likely a main sequence star with a mass of 0.18~$M_{\odot}$.
At least two eclipses have been observed per orbital cycle.
The longer of the two, between orbital phase 0.1--0.5 is most likely due to a companion wind.
The shorter eclipse, occurring between phase of 0.06--0.12 lasts for less than 10 minutes.

The discovery of PSR~J1717+4308A or M92A demonstrates the potential for FAST as an excellent probe of the GC pulsar population in the coming years,
which suggested that the total population of pulsars in M92 should be one of the highest in the FAST sky.
There are other 24 even richer clusters to be searched, including NGC~7078, NGC~7089 and Pal 2 (see Table 3 of Zhang et al.~(2016) for details).
A full census of the FAST GC pulsar sample will undoubtedly lead to interesting individual binary systems.

\acknowledgments
This work is supported by the Basic Science Center Project of the National Nature Science Foundation of China (NSFC) under Grant No. 11988101,
the National Nature Science Foundation of China (NSFC) under Grant No 11725313, No. 11803051, No. U1631237, No. 11703047, No. 11773041, No. 11690024, No. 11743002, and No. 11873067,
by National Key R\&D Program of China No. 2017YFA0402600,
and State Key Development Program for Basic Research No. 2015CB857100,
the Strategic Priority Research Program of the CAS Grant No. XDB23000000.
This work made use of data from the Five-hundred-meter Aperture Spherical radio Telescope (FAST).
FAST is a Chinese national mega-science facility, built and operated by the National Astronomical Observatories, Chinese Academy of Sciences (NAOC).
We appreciate all the people from FAST group for their support and assistance during the observations.
The National Radio Astronomy Observatory is a facility of the National Science Foundation operated under cooperative agreement by Associated Universities, Inc.
The Green Bank Observatory is a facility of the National Science Foundation (NSF) operated under cooperative agreement by Associated Universities, Inc.
ZP is supoorted by the CAS "Light of West China" Program.
DRL acknowledges support from the National Science Foundation awards AAG-1616042, OIA-1458952 and PHY-1430284 and from the Research Corporation for Scientific Advancement.
SMR is a CIFAR Fellow and is supported by the NSF Physics Frontiers Center award 1430284.
WF acknowledges the STEM Mountains of Excellence Graduate Fellowship.
WZ is supported by the CAS Pioneer Hundred Talents Program.
KL acknowledges the financial support by the European Research Council for the ERC Synergy Grant BlackHoleCam (Grant No. 610058),
the FAST FELLOWSHIP from Special Funding for Advanced Users budgeted and administrated by Center for Astronomical Mega-Science,
Chinese Academy of Sciences (CAMS),
and the CAS-MPG LEGACY funding "Low-Frequency Gravitational Wave Astronomy and Gravitational Physics in Space".

\end{document}